\documentstyle[preprint,prl,aps,psfig]{revtex}         

\newcounter{FAQ}
\begin{document}
\draft

\title{Two interacting particles at the metal-insulator transition}

\author{Andrzej Eilmes$^{1}$, Uwe Grimm$^{2}$, Rudolf A.\ 
  R\"{o}mer$^{2}$, and Michael Schreiber$^2$\\ $^1$Department of
  Computational Methods in Chemistry,\\ Jagiellonian University,
  Ingardena 3, 30-060 Krak\'{o}w, Poland\\ $^2$Institut f\"{u}r
  Physik, Technische Universit\"{a}t, D-09107 Chemnitz, Germany }

\date{$Revision: 1.15 $; compiled \today}
\maketitle

\begin{abstract}
  To investigate the influence of electronic interaction on the
  metal-insulator transition (MIT), we consider the Aubry-Andr\'{e}
  (or Harper) model which describes a quasiperiodic one-dimensional
  quantum system of non-interacting electrons and exhibits an MIT. For
  a two-particle system, we study the effect of a Hubbard interaction
  on the transition by means of the transfer-matrix method and
  finite-size scaling. In agreement with previous studies we find that
  the interaction localizes some states in the otherwise metallic
  phase of the system.  Nevertheless, the MIT remains unaffected by
  the interaction.  For a long-range interaction, many more states
  become localized for sufficiently large interaction strength and the
  MIT appears to shift towards smaller quasiperiodic potential
  strength.
\end{abstract}


%
%

\section{Introduction}
\label{sec-intro}

The physics of the metal-insulator transition (MIT) continues to be at
the center of current research activities. For two decades it has been
known from the scaling hypothesis of localization \cite{AALR79} that
generically a disorder-driven MIT \cite{A58} in a free electron system
only occurs in more than two spatial dimensions, whereas in one or two
dimensions an arbitrarily small disorder will localize the electronic
wave functions. The relevance of many-particle interactions for the
MIT is much less understood \cite{BK94,KDS96}. Here we consider the
perhaps simplest tractable model of an interacting system at the MIT.
Namely, we study the case of just two interacting particles (TIP) in a
particular one-dimensional (1D) quasiperiodic (QP) potential. For a
single particle (SP) this QP model exhibits an MIT as a function of
the non-random QP potential strength.
 
The problem of TIP in a 1D random potential, where the wave functions
are always localized such that there is no MIT, has already been
studied in much detail
\cite{S94,S96,I95,FMPW95,OWM96,WMPF95,RS97,SK97,PS97}. It was argued
that a Hubbard onsite interaction $U$ dramatically reduces the
localization of TIP pair states in comparison with non-interacting and
unpaired particles. In particular, Shepelyansky \cite{S94,S96}
proposed an enhancement of the TIP localization length $\lambda_2$
independent of the statistics of the particles and of the sign of the
interaction such that
\begin{equation}
  \lambda_2(U) \approx U^2 \frac{\lambda_1^\kappa}{32}
\label{eq-shep}
\end{equation}
in the band center with $\kappa=2$. Here, $\lambda_1$ is the SP
localization length in 1D \cite{CKM81} and $U$ is given in units of
the nearest-neighbor hopping strength.
Microscopic support for the delocalization was given afterwards by
Frahm {\em et al}.\ \cite{FMPW95}, who observed a behavior $\lambda_2
\sim \lambda_1^{1.65}$ in a numerical investigation employing the
transfer-matrix method (TMM). Other direct numerical approaches to the
TIP problem have been based on the time evolution of wave packets
\cite{S94,BGKTMV98}, exact diagonalization \cite{WMPF95}, Green
function approaches \cite{OWM96,SK97,LRS98}, and TMM
\cite{RS97,HZKM98}.  In these investigations an enhancement of
$\lambda_2$ compared to $\lambda_1$ has usually been found, but the
quantitative results tend to differ both from the analytical
prediction (\ref{eq-shep}), and from each other. 

Two of us \cite{RS97} recently studied the TIP problem by TMM but at
larger system sizes $M$ than Ref.\ \onlinecite{FMPW95} and found that
(i) the enhancement $\lambda_2/\lambda_1$ decreases with increasing
$M$, (ii) the behavior of $\lambda_2$ for $U=0$ is equal to
$\lambda_1$ in the limit $M\rightarrow\infty$ only, and (iii) for
$U\neq 0$ the enhancement $\lambda_2/\lambda_1$ also vanishes
completely in this limit. This raises serious questions about the
validity of the TMM approach to TIP and in fact it has been argued
very recently \cite{OS98}, that the TMM approach of Ref.\ 
\onlinecite{FMPW95,RS97} may systematically underestimate the
localization length of a pair state, since it automatically measures a
mixture of localization lengths originating also from unpaired states.
Thus in this work, we will use the TIP-TMM not as a tool to extract
information about the pair states only, but rather aim at describing
the general influence of the presence of one particle onto the
transport properties of the other.

At present, it seems well-established by Green function methods
\cite{OWM96,SK97,LRS98} that an enhancement $\lambda_2 > \lambda_1$
exists, although the validity of Eq.\ (\ref{eq-shep}) is still under
debate: the values of the exponent $\kappa$ obtained by numerical
methods \cite{S94,FMPW95,OWM96,WMPF95,RS97,SK97,BGKTMV98,LRS98,HZKM98}
range from $1$ to $2$.  In spite of these numerical differences, we
nevertheless believe that the TIP approach can give meaningful
insights into the interplay of disorder and interaction \cite{LRS98}.
In particular, the effects of interaction on the disorder-driven
Anderson transition should be quite interesting already for TIP.
However, as mentioned above, the disorder-driven MIT requires more
than two spatial dimensions and so the numerical efforts are close to
being prohibitive when including interactions.

Fortunately, the QP --- and thus fully deterministic ---
Aubry-Andr\'{e} (AA) model \cite{AA80} exhibits an MIT even in 1D, in
dependence on the strength of the quasiperiodic potential.  This model
is closely related to the problem of a SP on a 2D lattice in a
magnetic field in which context it is also known as the Harper model
\cite{H55}. At the MIT, the spectrum exhibits the famous Hofstadter
butterfly shape \cite{H76}, and the spectral and localization
properties have been studied in great detail \cite{K83}.  In the
mathematical literature, the same model is also known and studied as
the almost-Mathieu equation \cite{BLT83}.

For this 1D model, we can use the TIP approach in a straightforward
way in order to investigate the effect of the interaction on the
transition.  Previous studies based on perturbative expansions in $U$
and numerical computations of participation numbers in the AA model
\cite{BBJS96} concluded that interaction can lead to the appearance of
localized states in the metallic regime for TIP. However, although
participation numbers are a useful tool for characterizing
localization properties of states, they may give ambiguous results: in
some cases, states which are extended or critical may appear to be
more localized and vice versa \cite{ERS98}.  Moreover, for interacting
particles the generalization of localization criteria like the
participation number is not straightforward \cite{VES97}. Thus in this
work we concentrate on direct calculations of the TIP localization
length in the AA model using the TMM for finite system sizes. In
addition to the onsite interaction, we also consider a long-range
interaction. Employing the finite-size-scaling (FSS) approach
\cite{MK83}, we then construct scaling curves from which we deduce the
localization properties of the infinite system. We find that within
the accuracy of our results, the critical behavior is not affected by
the interactions. But it seems that the long-range interaction shifts
the critical QP potential strengths towards smaller values, thus
giving a tendency towards localization.

The paper is organized as follows. In section \ref{sec-tipaa}, we
define the TIP version of the AA model and introduce our notations.
Section \ref{sec-tmm} reviews the power-series variant of the TMM, and
the concepts of FSS. In section \ref{sec-beta} we explain the use of a
phase-shift parameter in the QP potential for reducing statistical
fluctuations in the localization length data.  Results obtained from
FSS of the localization lengths for Hubbard and long-range
interactions at energy $E=0$ are presented in section
\ref{sec-bandcenter}. In section \ref{sec-band}, we show the
localization properties of all states of the spectrum. We summarize
and conclude in section \ref{sec-con}.

%
%

\section{The TIP version of the Aubry-Andr\'{e} model}
\label{sec-tipaa}

The Schr\"{o}dinger equation for the SP AA model is given as
\begin{eqnarray} 
\label{eq-aa}
  \phi_{n+1} = ( E - \mu_n ) \phi_{n} - \phi_{n-1}.
\end{eqnarray}
Here $\phi_n$ is a SP wave function, $\mu_n \equiv 2 \mu \cos( \alpha
n + \beta)$ is the QP AA onsite potential of strength $\mu$ with
$\alpha/2 \pi$ an irrational number, which we have chosen as the
inverse of the golden mean $\alpha/2 \pi= (\sqrt{5} -1)/2$, and
$\beta$ is an arbitrary phase shift.  We remark that $\alpha/2\pi$ may
be approximated by the ratio of successive Fibonacci numbers $1, 2, 3,
5, 8, 13,\ldots$.  In Fig.\ \ref{figaa1n} we show typical data for the
SP localization length $\lambda_1$ obtained by TMM for various system
sizes given by some of the Fibonacci numbers \cite{evenodd}. In
agreement with previous studies \cite{AA80}, this figure suggests
already that the MIT occurs ar $\mu = 1$. Of course, further analysis
like FSS would be necessary for a comprehensive study of this MIT.
Here we note that in contradistinction to the MIT in the usual
Anderson model with onsite {\em random} potential disorder, in the AA
model {\em all} states are either extended ($\mu<1$), critical
($\mu=1$), or localized ($\mu>1$), and thus no mobility edge, i.e., no
MIT in dependence {\em on energy} exists.

In principle, there are many possibilities to extend the SP
Schr\"{o}dinger equation to TIP. In order to be most compatible with
the TIP approach of Shepelyansky \cite{S94}, we will consider a TIP
Hamiltonian with an additional QP onsite potential on
a chain of length $M$ given as
\begin{eqnarray} \label{hamilt}
H & = &
   \sum_{n=1}^{M}
   (c^{\dag}_{n+1 }c^{ }_{n } + h.c.)
   +\sum_{n=1}^{M} \sum_{m=1}^{M} U_{n,m} 
   c^{\dag}_{n} c^{ }_{n} 
   c^{\dag}_{m} c^{ }_{m},
   \nonumber \\ & & \mbox{}
   +\sum_{n=1}^{M} \mu_n c^{\dag}_{n} c^{ }_{n}
\end{eqnarray}
where $c_{n}^{\dag}$ and $c^{ }_{n}$ are the creation and annihilation
operators for the electron at site $n$ and we assume that the TIP have
different spins. $U_{n,m}$ denotes the interaction between particles:
$U_{n,m}=U \delta_{nm}$ for Hubbard onsite interaction or
$U_{n,m}=U/(|n-m|+1)$ for long-range interaction.

%
%

\section{The transfer-matrix approach to TIP}
\label{sec-tmm}

The TIP Schr\"odinger equation reads
\begin{eqnarray} \label{eq-tip}
  \psi_{n+1,m} & = & [ E - U_{n,m} - \mu_n - \mu_m ] \psi_{nm} 
  \nonumber \\ & & -
  \psi_{n,m+1} -\psi_{n,m-1} -\psi_{n-1,m},
\end{eqnarray}
with $\psi_{n,m}$ a TIP wave function which at $U=0$ may be written as
a product of SP wave functions $\phi_n$ and $\phi_m$.  We can rewrite
Eq.~(\ref{eq-tip}) in the TMM form similar to a 2D Anderson model on
an $M \times M$ lattice as
$( \psi_{n+1}, \psi_{n} )^{T} = T_n (\psi_{n}, \psi_{n-1} )^{T}$ with
the symplectic transfer matrix
\begin{equation}
  T_n= \left(
 \begin{array}{cc}
   E \openone - \chi_n - H_{\perp} \quad & - \openone \\ \openone &
   {\bf 0}
 \end{array}
\right),
\end{equation}
describing the evolution of the wave vectors for the first ($n$)
particle (corresponding to the longitudinal direction in the 2D SP TMM
approach).  Here $\psi_n= (\psi_{n,1}, \ldots, \psi_{n,m}, \ldots,
\psi_{n,M})$ is the wave vector of slice $n$, $H_{\perp}$ is the SP
hopping term for the second ($m$) particle (corresponding to the
transverse direction) and $(\chi_n)_{i,m}= [ \mu_n + \mu_m + U_{n,m} ]
\delta_{i,m}$ codes the QP potential and the interaction
\cite{FMPW95}. Note that in this approach the symmetry of the wave
function remains unspecified and we cannot distinguish between boson
and fermion statistics.

The evolution of the state is determined by the matrix product
$\tau_N= \prod_{n=1}^{N} T_n$ and we have
\begin{equation}
  \left( \begin{array}{l} \psi_{N+1} \\ \psi_{N} \end{array} \right) =
  \tau_N \left( \begin{array}{l} \psi_{1} \\ \psi_{0} \end{array}
  \right).
\label{eq-globtmm}
\end{equation}
Usually, one studies a quasi-1D system of size $M \times N$ with $M\ll
N$. However, in the present problem, both directions are restricted to
$n,m \leq M$ and iterating Eq.~(\ref{eq-globtmm}) only $N=M$ times
will not give convergence.  Frahm {\em et al}.\ \cite{FMPW95} have
solved this problem in their TMM study by exploiting the Hermiticity
of the product matrix $Q_M= \tau^\dagger_M \tau_M$: Continuing the
iteration (\ref{eq-globtmm}) with $\tau^\dagger_M$, then with
$\tau_M$, and so on, until convergence is achieved, yields the
eigenvalues $\exp[-2 M \gamma_i]$ of $Q_M$. This is the well-known
power method for the diagonalization of Hermitian matrices
\cite{HJ85}. The smallest positive Lyapunov exponent $\gamma_{\rm
  min}$ determines the slowest possible decay of the wave function and
thus the largest localization length $\lambda_{\rm max}= 1/\gamma_{\rm
  min}$ for given energy $E$ and phase shift $\beta$. We now {\em
  define} the localization length $\lambda$ of the two-particle wave
function $\psi_{n,m}$ as $\lambda_{\rm max}$ of the transfer matrix
problem (\ref{eq-globtmm}) and expect it to reflect the influence of
the particle interaction.

According to the one-parameter scaling hypothesis \cite{AALR79}, which
has been verified with very high accuracy for random potentials
$\mu_n$ \cite{MK83}, the reduced localization lengths $\lambda(M)/M$
scale onto a single scaling curve, i.e.,
\begin{equation}
  \lambda(M)/M = f(\xi /M).
\label{eq-fss}
\end{equation}
For the AA model considered here, we are not aware of any previous FSS
study. Indeed, it is not a priori obvious that one-parameter FSS
should be valid for the AA model. At least for a given single phase
shift $\beta$, it is clear from Fig.\ \ref{figaa1n} that we need to go
to rather large system sizes in order to suppress the fluctuations
around $\mu=1$ and to be able to use the FSS approach.  However, as we
will explain in the next section, we may use different values of
$\beta$ as being analogous to the different disorder realizations in
the Anderson model.  As usual, we may then determine the
finite-size-scaling (FSS) function $f$ and the values of the scaling
parameter $\xi$ by a least-squares fit \cite{MK83}.

%
%

\section{Averaging over different \protect\boldmath{$\beta$}}
\label{sec-beta}

The localization length calculated for given system size and QP
potential $\mu$ depends significantly on the $\beta$ value as shown
for SP in Fig.\ \ref{figbeta}. 
This means that the decay length varies depending on the phase shift
of the potential along the chain. One may expect that the chain length
$M$ will also influence the results by changing relative phases of the
potential at the ends.

Therefore we have restricted our calculations to the chain lengths
given by the Fibonacci numbers mentioned in section \ref{sec-tipaa},
because for our choice of $\alpha$, this assures that the phase
difference of the potential at both ends of the chain will be similar,
i.e., approaching zero with increasing $M$.  We note that our
numerical results presented in the next section do not change
significantly, when we alternatively use the rational approximants for
$\alpha/2\pi$ instead of the irrational number defined in section
\ref{sec-tipaa}.

Still, the dependence of $\lambda_1$ on the system size $M$ for a
given value of $\beta$ shows much structure which makes simple
extrapolations towards the infinite system or FSS impossible.  This
dependence is also responsible for fluctuations of the SP $\lambda_1$
close to $\mu=1$ which are visible as peaks in Fig.\ \ref{figaa1n} for
small Fibonacci number $M$ \cite{evenodd}. Only for very large $M$,
the fluctuations become small.  On the other hand, finite systems with
different values of $\beta$ may be viewed as different parts cut out
of the infinite QP model.  This then suggests that we may reduce the
fluctuation effects by averaging over many such small pieces or,
equivalently, many different values of randomly chosen $\beta$. Thus
different $\beta$ values are analogous to different disorder
configurations used in the Anderson Hamiltonian.  Fig.\ \ref{figaan}
presents such an average over $1000$ $\beta$ values for the SP
localization length.  As expected, the fluctuations visible in Fig.\ 
\ref{figaa1n} disappear even for small systems and extrapolations to
large $M$ and FSS are now possible.


%
%

\section{Localization properties at \protect\boldmath{$E=0$}}
\label{sec-bandcenter}

We now turn our attention to the problem of TIP and study the effects
of interaction on the localization lengths obtained by TMM,
restricting ourselves to $E=0$. To this end, we have computed the
localization lengths for $6$ system sizes $M= 8, 13, 21, 34, 55$, and
$89$, for $80$ QP potential strengths $\mu$ ranging from $0.56$ to
$4$, and for $6$ interaction strengths $U= 0, 0.5, 1, 1.5, 2$, and
$10$, with onsite and with long-range interaction.  Typically, for
each such triplet of parameters $(M,\mu,U)$ we averaged over at least
$1000$ different $\beta$ realizations.  We note that as for the case
of TIP in a random potential \cite{LRS98}, attractive and repulsive
interaction strengths give the same results at $E=0$ and we can thus
restrict ourselves to $U \geq 0$ here.

\subsection{Hubbard interaction}
\label{subsec-bch}

Figs.\ \ref{fig-fssU0} and \ref{fig-fssU1} show the FSS results for
$\beta$-averaged data at energy $E=0$ for onsite interaction strength
$U=0$ and $U=1$. As can be seen, the coalescence of data for various
values of $\mu$ is not perfect and in fact certainly worse than, e.g.,
for onsite random disorder \cite{MK83}. This is especially visible on
the extended side $\mu < 1$. Nevertheless, the figures clearly show
the existence of two branches of the scaling curve as in the 3D
Anderson model \cite{MK83}. This indicates, in agreement with the
above considerations for the SP AA model, the presence of localized
states for $\mu>1$ and extended states for $\mu<1$. The MIT appears at
a critical QP potential $\mu_c$ which is close to $1$.  The values
determined from the FSS procedure are $\mu_c= 1.01 \pm 0.02$ for $U=0$
and $\mu_c=1.04 \pm 0.04$ for $U=1$. Corresponding FSS plots for $U=
0.5, 1.5, 2$ and $10$ all consistently give $\mu_c\approx 1$.  We
attribute the small deviations from the critical value $\mu_c=1$ of
the SP case to the fluctuations in the data.

Thus the MIT does not get shifted by the Hubbard interaction and the
transport properties of one particle in the presence of another remain
unchanged. On the metallic side of the transition ($\mu<1$) this is
immediately clear: the interaction is supposed to localize ${\cal
  O}(M)$ TIP states out of the ${\cal O}(M^2)$ states in the
unsymmetrized Hilbert space \cite{BBJS96}. The TMM inherently measures
the longest localization length and thus simply misses the few shorter
localization lengths induced by the interaction. On the localized
side, however, we could expect the interaction to delocalize these TIP
states which might be visible even by TMM. However, as discussed in
section \ref{sec-intro}, this effect is not present in the TIP-TMM or
at least too small to be visible \cite{OS98}.

The scaling parameters $\xi$ obtained by FSS according to Eq.\ 
(\ref{eq-fss}) are expected to diverge at the transition as $\xi \sim
|\mu - \mu_c|^{-\nu}$ with the critical exponent $\nu$. In Fig.\ 
\ref{figscp} we show the dependence of $\xi$ on the QP strength $\mu$. The
divergence at $\mu_c \approx 1$ is clearly visible. A power-law fit
gives $\nu = 0.8 \pm 0.2$ both for $U=0$ and $U=1$. The large error of
the estimate is due to the fluctuations in the data near the critical
point \cite{MK83}.  Furthermore, in the localized regime of the SP AA
model is has been shown that \cite{AA80}
\begin{equation}
  \lambda_1 \sim 1/\ln(1+|\mu - \mu_c|),
\label{eq-ixi}
\end{equation}  
which yields $\nu=1$ by expansion around $\mu_c$.  In order to check
whether this equation holds also for TIP we examined the dependence of
$1/\xi$ on $\ln{(1+|\mu - \mu_c|)}$. The results are displayed in
Fig.\ \ref{figlocl}. The slope of the best fit line is $1.00 \pm 0.02$
for $U=0$ and $1.01 \pm 0.03$ for $U=1$.  This suggests that onsite
interaction does not change the critical behavior at the MIT.

\subsection{Long-range interaction}
\label{subsec-bclr}

We now consider the long-range interaction defined in section
\ref{sec-tipaa}.  The FSS plot for $U=1$ in Fig.\ \ref{fig-fssLR} is
qualitatively the same as for Hubbard interaction.  We find localized
states for $\mu \gg 1$ and extended states for $\mu \ll 1$. In Fig.\ 
\ref{figscp}, we have included the variation of the scaling parameter
$\xi$ with $\mu$ for this case. The divergence of $\xi$ occurs at
$\mu_c=0.92\pm 0.04$ indicating that the MIT has been shifted towards
smaller values of the QP potential strength $\mu$. FSS plots for
$U=0.5, 1.5$ and $2$ suggest that this shift becomes somewhat more
pronounced for larger $U$, decreasing to $\mu_c\approx 0.9$ for $U=2$.

This behavior may be rationalized by keeping in mind that for a
long-range interaction, contrary to the case of Hubbard interaction,
{\em all}\/ states will eventually feel the interaction-induced
tendency towards localization on the extended side of the MIT, as we
will show for small systems in the next section. Thus even the most
delocalized states at $E=0$ will become more localized for
sufficiently large $U$.  However, in order to answer the question
whether long-range interaction indeed shifts the MIT towards weaker QP
potential strength, additional calculations with still higher accuracy
would be necessary.  These require, however, a prohibitive numerical
effort when using the present power-series method.

The critical exponent for $U=1$ calculated as in section
\ref{subsec-bch} is $\nu = 1.0 \pm 0.2$ and the respective slope in
Fig.\ \ref{figlocl} is $0.97 \pm 0.03$. These values are compatible
with our results for onsite interaction within the error limits.
Therefore, the critical behavior is similar to the SP case and onsite
interaction.
 
%
%

\section{Energy dependence of the localization properties}
\label{sec-band}

In the previous section, we have rationalized the persistence of the
MIT in the presence of interactions by assuming that on the extended
side the onsite interaction localizes a small number of states leaving
the rest unaffected. To further examine this effect with TMM we
calculate the dependence of $\lambda$ on the energy $E$ for a single
value of $\beta$ and a small system size.

\subsection{Hubbard interaction}
\label{subsec-bh}

Fig.\ \ref{fig-EdepExt} presents results for the inverse localization
length $\lambda^{-1}$ obtained by TMM on the metallic side. Also shown
are the values of the eigenenergies $E_i$.
The TMM accurately shows that transport at energies not corresponding
to eigenstates is suppressed, because the incoming wave function
decays exponentially.  On the other hand, $\lambda^{-1}(E)$ decreases
rapidly towards zero when $E$ is approaching an eigenvalue $E_i$ as
shown in Fig.\ \ref{fig-EdepExt}.  This has also been observed in the
SP case \cite{AA80}.  For $U=0$, we find a few cases where
$\lambda^{-1}$ remains large even at the energy of an eigenstate.
From an analysis of the corresponding wave functions, we can identify
these states with boundary states where the particles are localized
close to the ends of the finite chains.

The comparison of the plots for $U=0$ and $U=1$ shows that, while the
energy of most states changes only slightly, there are a few states
which move to significantly larger energies. Their localization
lengths are apparently much shorter. The calculation of
$\beta$-averaged decay lengths for different $M$ shows that states at
the verge of the spectrum at $E=4.6$ remain extended for $U=1$ while
the states at $E=5.3$ are localized. There are also some states within
the main part of the spectrum which shift to higher energies.  Some of
them are visible in Fig.\ \ref{fig-EdepExt} as they enter the energy
gaps.  For sufficiently strong interaction $U=8$ there are $13$
localized states which split off the remaining spectrum. The
calculations for other system sizes support the conclusion that the
interaction localizes $M$ out of $M^2$ states for system size $M$.
These states correspond to both particles residing on the same site
and interacting via the Hubbard $U$. The other states remain extended
and do not change their energy significantly.

In the localized regime ($\mu > 1$) the interaction has a similar
effect, i.e., for sufficiently large $U$ it shifts $M$ states above
the main part of the spectrum and increases their localization; the
remaining unshifted states also stay localized. These results are in
agreement with those of Ref.\ \cite{BBJS96}, when we keep in mind that
our numerical method does not allow us to see accurately an eventual
delocalization at intermediate $U$.

\subsection{Long-range interaction}
\label{subsec-blr}

Fig.\ \ref{fig-EdepExtLR} presents respective results obtained for
long-range interaction. Again, the interaction shifts states to higher
energies and shortens their decay lengths. However, as the particles
feel the interaction at any separation, {\em all}\/ states change
their energy in agreement with section \ref{subsec-bclr}. This is
especially pronounced in Fig.\ \ref{fig-EdepExtLR} for $U=10$.  The
most prominent shift is the change at the high energy part of the
spectrum. For extremely large interaction, e.g.\ $U=1000$, the
spectrum splits into $M$ groups of states reflecting the number of
sites at which two particles may reside at given separation, i.e., for
system size $M$ there are $M$ states for separation $n-m=0$, and
$2M-2$ states for separation $|n-m|=1$, and so on.

%
%

\section{Conclusions}
\label{sec-con}

We have demonstrated that it is possible to perform FSS for a system
with two interacting particles in a 1D QP Aubry-Andr\'{e} (or Harper)
potential.  We find two branches in the FSS curves which correspond to
localized and extended behavior.  The roughness of the FSS plot is
probably an effect of small system sizes and insufficient averaging
and should disappear for larger systems, requiring, however, much
larger computational effort. On the other hand it may be that
one-parameter scaling is not strictly valid in this QP model as
evidenced by the results for the even chain lengths $M=34$ and $144$
\cite{evenodd}.  Nevertheless, even in this case the presence of
localized and extended branches as in Figs.  \ref{fig-fssU0},
\ref{fig-fssU1}, \ref{fig-fssLR} indicates the existence of an MIT.

The FSS results for energy $E=0$ show that the MIT exists in these TIP
systems both for the non-interacting and the interacting case.  The
transition point $\mu_c$ does not depend on the Hubbard interaction
strength $U$ and is located at QP potential strength $\mu_c \approx
1$.  However, a large enough long-range interaction shifts $\mu_c$
towards smaller QP strength. Within the numerical accuracy of our
data, the critical behaviour of the localization length is not
affected by the Hubbard and the long-range interaction.

The dependence of the decay length on the energy as calculated by TMM
confirms the results obtained by other methods \cite{AA80} that a
large enough interaction localizes pair states simultaneously
increasing their energy and leaves the rest of the states almost
unaffected.

In closing we remark that our results may also be viewed independently
of the TIP problem, by noting that the present problem of two
particles in a 1D QP potential may also be seen as SP problem in a
particular realization of a 2D QP potential.  Similar systems have
been investigated previously, e.g., in Ref.\ \cite{GGS98} within the
Landauer approach.

\acknowledgements

We thank K.\ Frahm and J.-L. Pichard for a clarifying discussion on
the power series method. A.E.\ expresses his gratitude to the
Foundation for Polish Science for a fellowship.  This work has been
supported by the Deutsche Forschungsgemeinschaft (SFB 393).

%
%

%
%
\newcommand{\figwidth}{0.9\columnwidth}

\begin{figure}
  \centerline{\psfig{figure=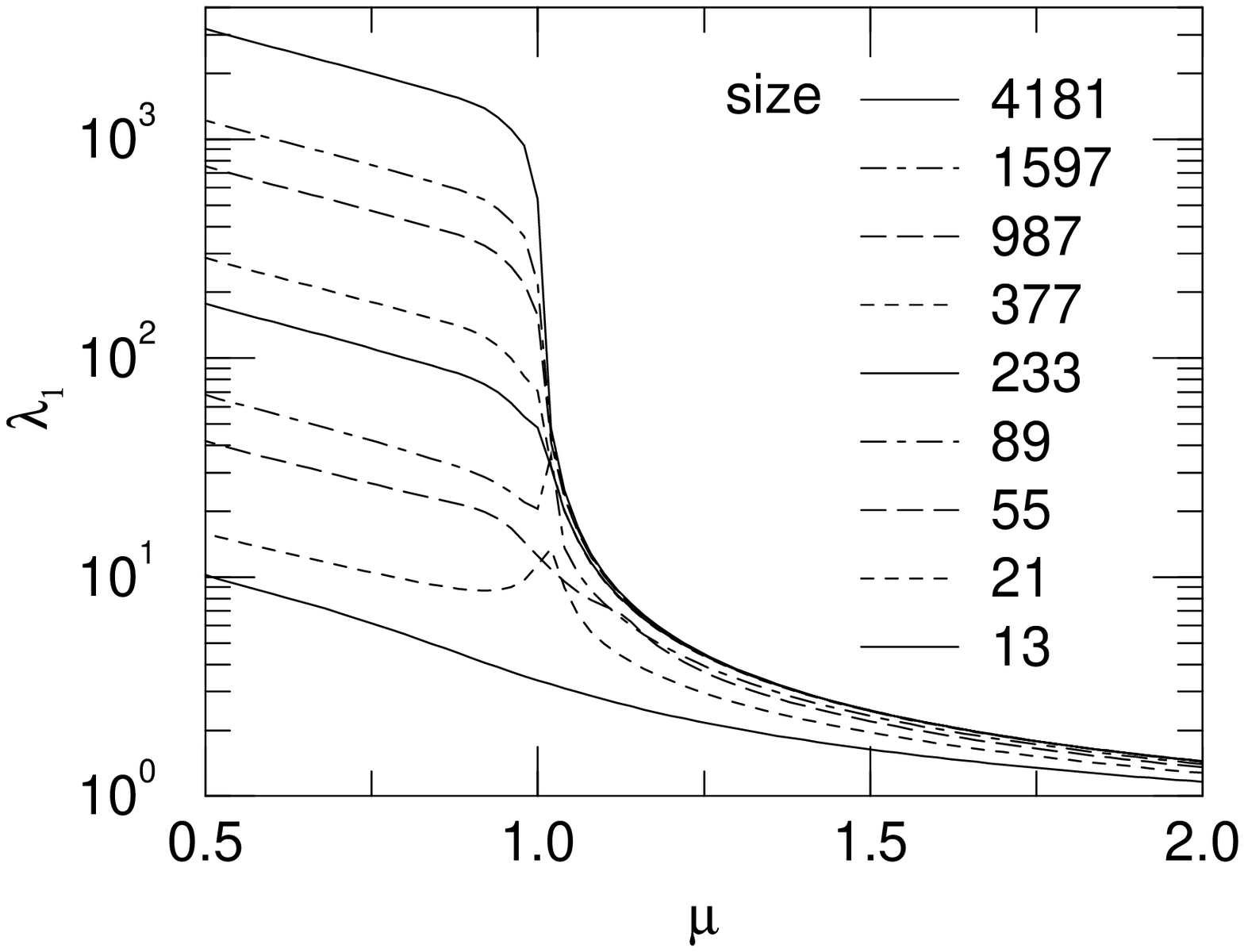,width=\figwidth}}
  \caption{
    Localization length $\lambda_1$ for the SP Aubry-Andr\'e model as a
    function of {QP} potential strength ${\mu}$ for $E=0$ and
    $\beta=\protect\sqrt{2}$ with system size increasing from bottom
    to top. }
\label{figaa1n}
\end{figure}

\begin{figure}
  \centerline{\psfig{figure=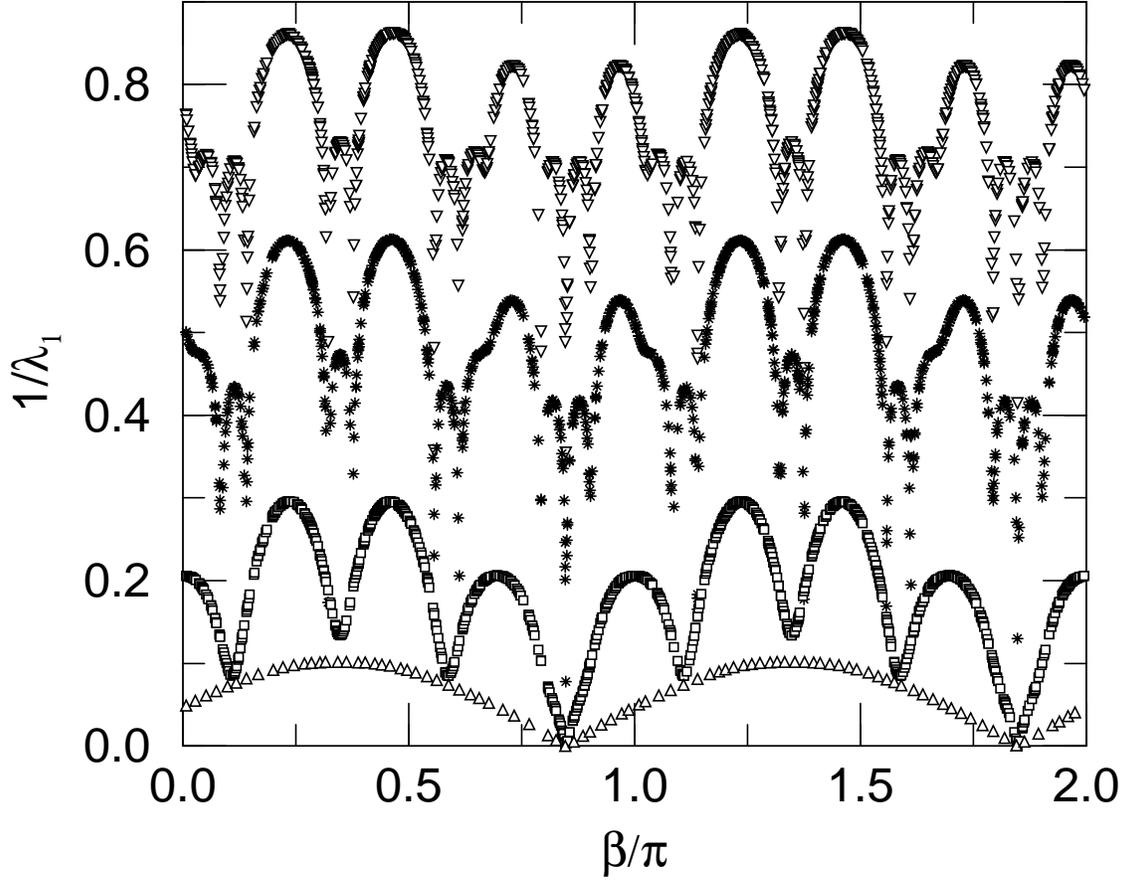,width=\figwidth}}
  \caption{
    Inverse of the localization length $\lambda_1$ for the SP
    Aubry-Andr\'e model as a function of phase shift $\beta$ for $E=0$
    and $M=13$. Different symbols indicate QP potential strength
    $\mu=2$ ($\nabla$), 1.5 ($*$), 1 ($\Box$), and 0.5 ($\triangle$).
    }
\label{figbeta}
\end{figure}

\begin{figure}
  \centerline{\psfig{figure=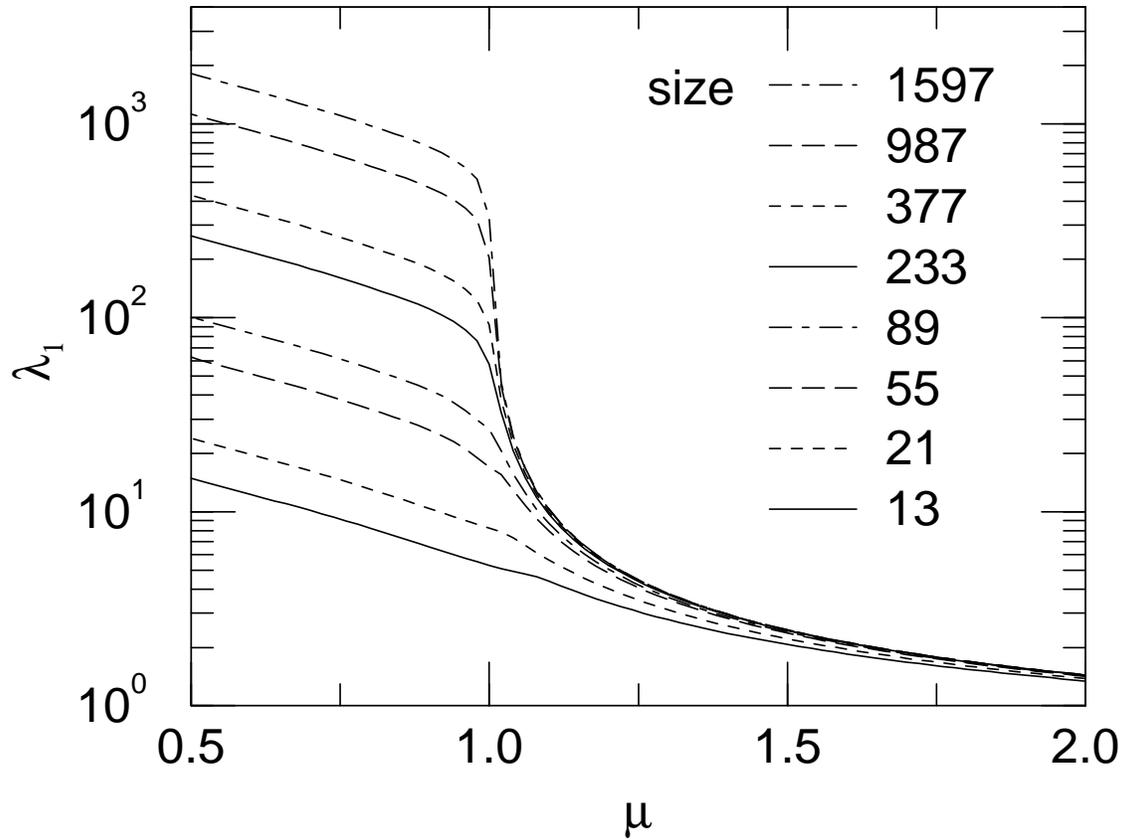,width=\figwidth}}
  \caption{
    {Localization} length $\lambda_1$ for the SP Aubry-Andr\'e model as
    a function of {QP} potential strength ${\mu}$ for $E=0$, averaged
    over $1000$ $\beta$-values. The system size is increasing from
    bottom to top. Note the MIT at ${\mu}=1$. }
\label{figaan}
\end{figure}

\begin{figure}
  \centerline{\psfig{figure=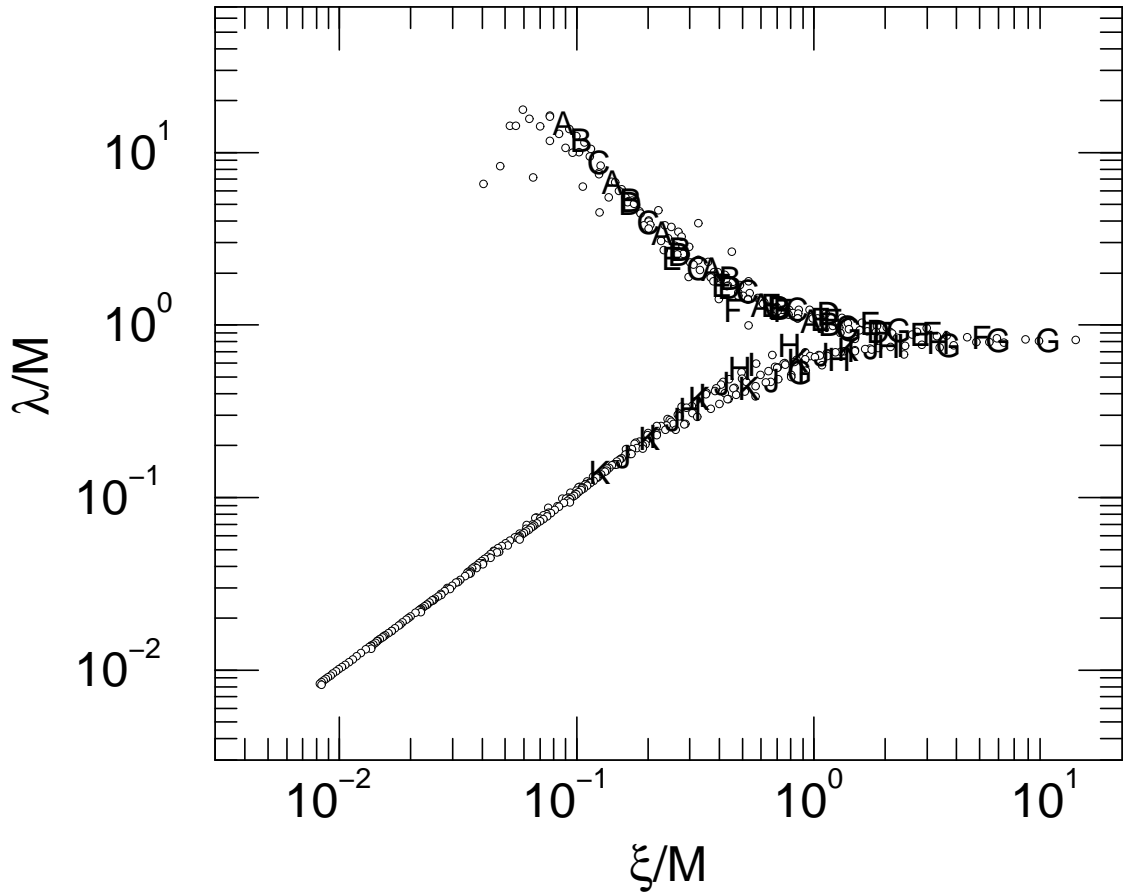,width=\figwidth}}
  \caption{
    Scaling function (\protect\ref{eq-fss}) for $U=0$, $E=0$ and
    various ${\mu}$. Data for $\mu=0.9$, $0.92$, $0.94$, $0.96$,
    $0.98$, $1.0, 1.02$, $1.04$, $1.06$, $1.08$, and $1.1$ are marked
    with characters A, B, \dots, K, respectively.}
\label{fig-fssU0}
\end{figure}

\begin{figure}
  \centerline{\psfig{figure=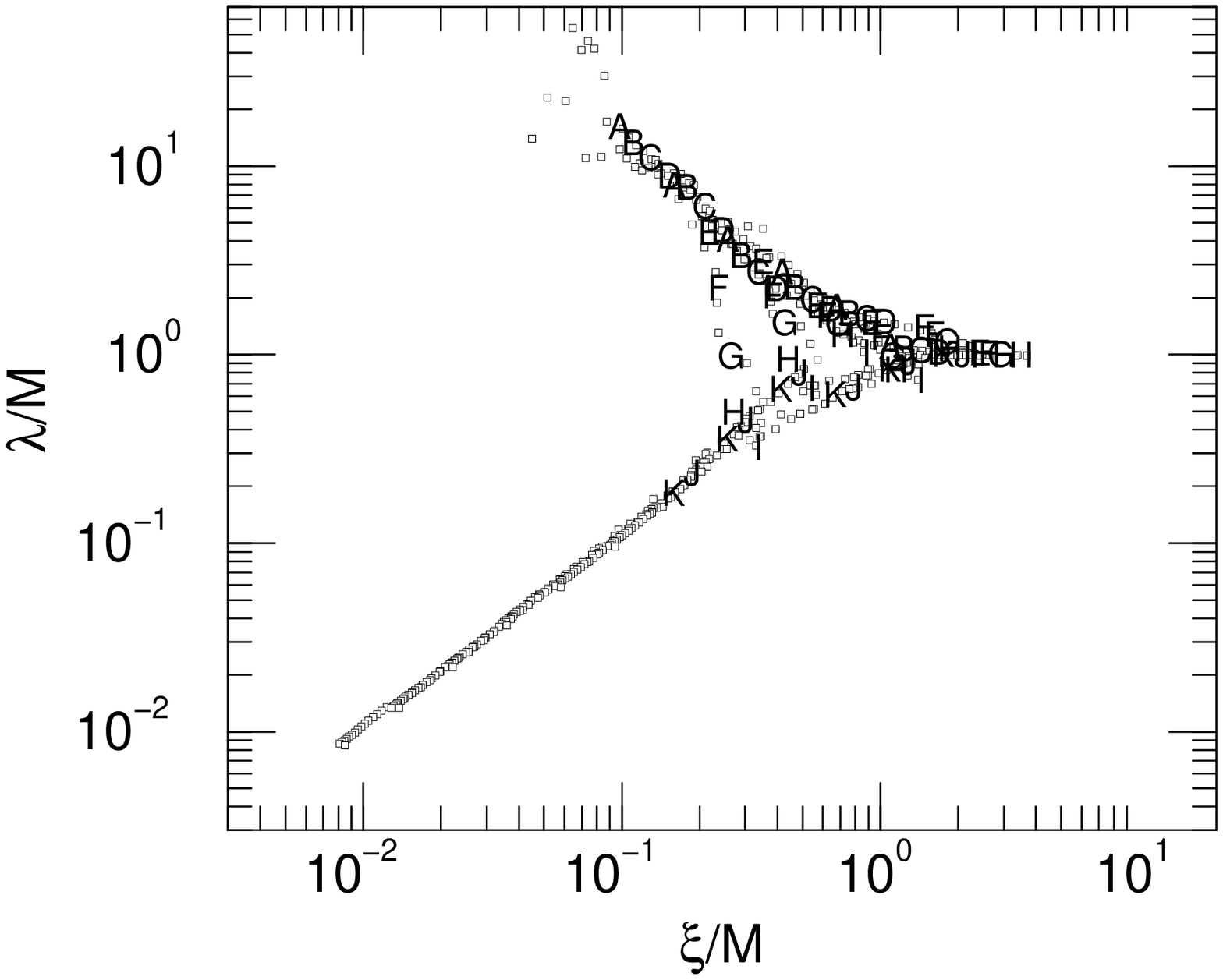,width=\figwidth}}
  \caption{
    Scaling function (\protect\ref{eq-fss}) for ${U}=1$, $E=0$ and
    various ${\mu}$. The characters are chosen as in Fig.\ 
    \ref{fig-fssU0}.}
\label{fig-fssU1}
\end{figure}

\begin{figure}
  \centerline{\psfig{figure=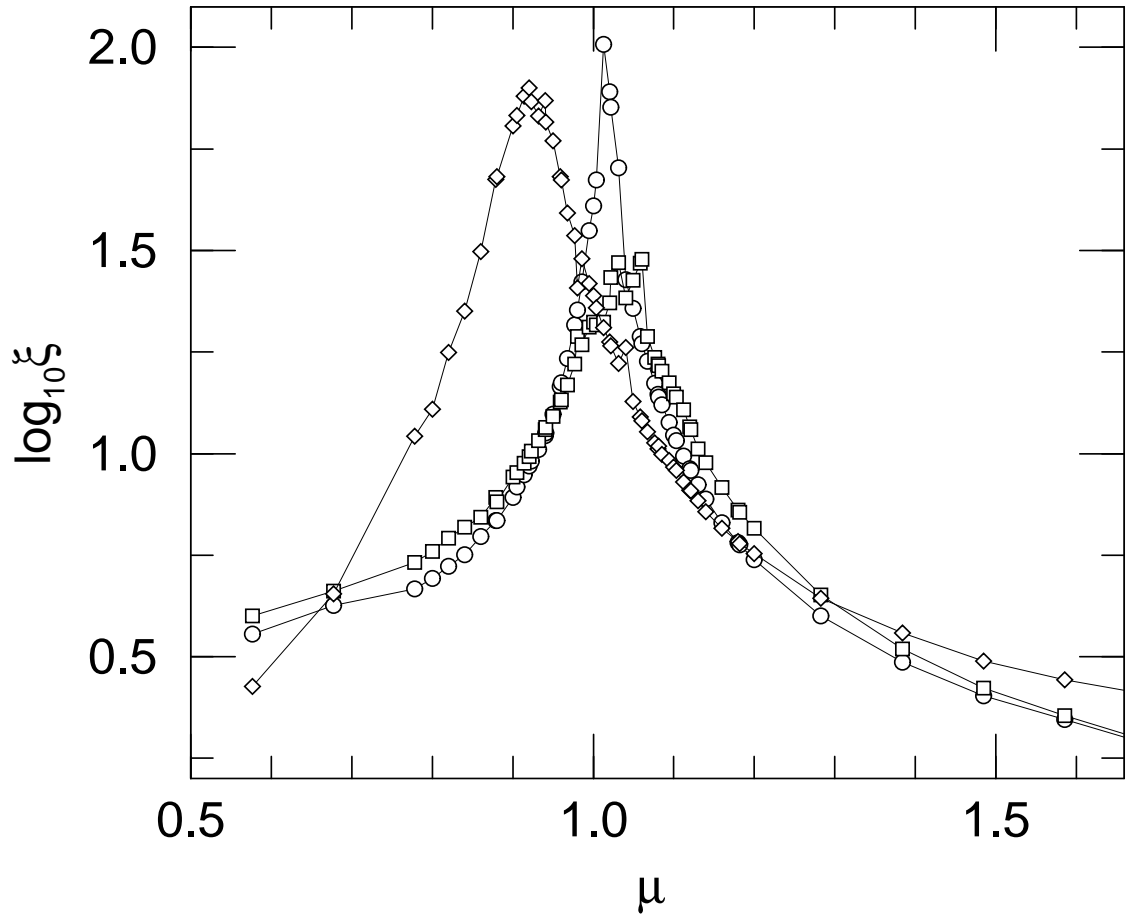,width=\figwidth}}
  \caption{
    Scaling parameter $\xi$ as a function of QP potential strength
    $\mu$ for $U=0$ ($\circ$), for onsite interaction with $U=1$
    ($\Box$), and for long-range interaction with $U=1$ ($\Diamond$).}
\label{figscp}
\end{figure}

\begin{figure}
  \centerline{\psfig{figure=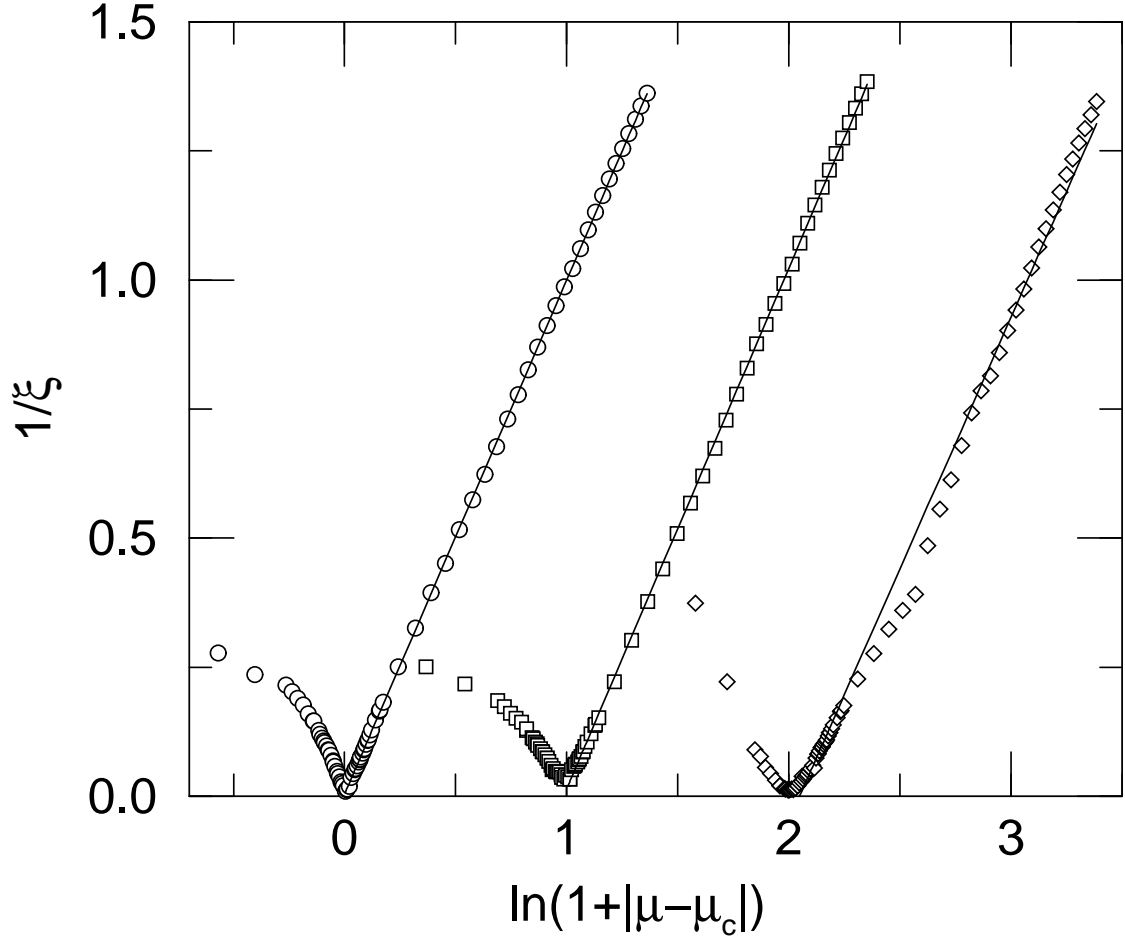,width=\figwidth}}
  \caption{
    Inverse scaling parameter $1/\xi$ as a function of QP potential
    strength $\mu$ as in Eq.\ (\protect\ref{eq-ixi}) for $U=0$
    ($\circ$), for Hubbard interaction with $U=1$ ($\Box$), and for
    long-range interaction with $U=1$ ($\Diamond$), consecutively
    shifted by 1 for clarity. The lines indicate linear regression
    fits to the data in the localized regime.}
\label{figlocl}
\end{figure}

\begin{figure}
  \centerline{\psfig{figure=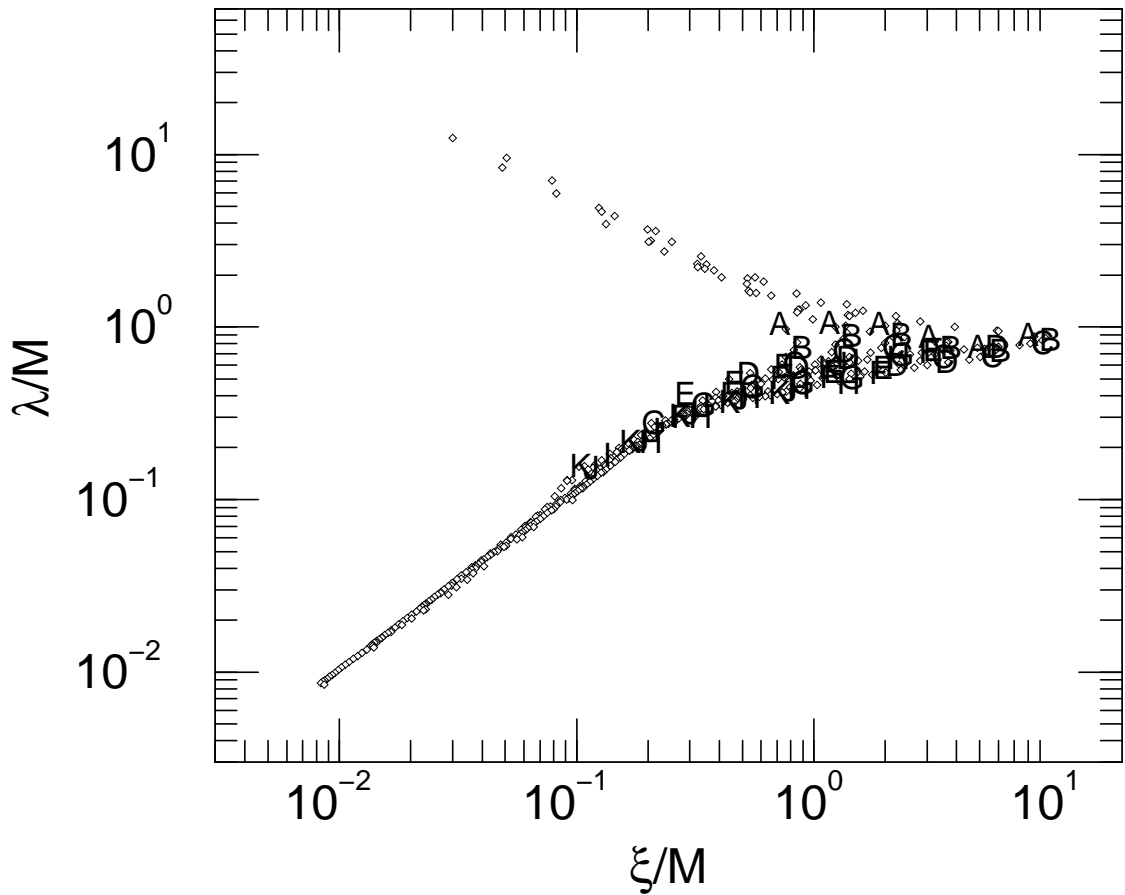,width=\figwidth}}
  \caption{
    Scaling function (\protect\ref{eq-fss}) for long-range interaction
    ${U}=1$, $E=0$ and various ${\mu}$. The characters are chosen as
    in Fig.\ \ref{fig-fssU0}.}
\label{fig-fssLR}
\end{figure}

\begin{figure}
  \centerline{\psfig{figure=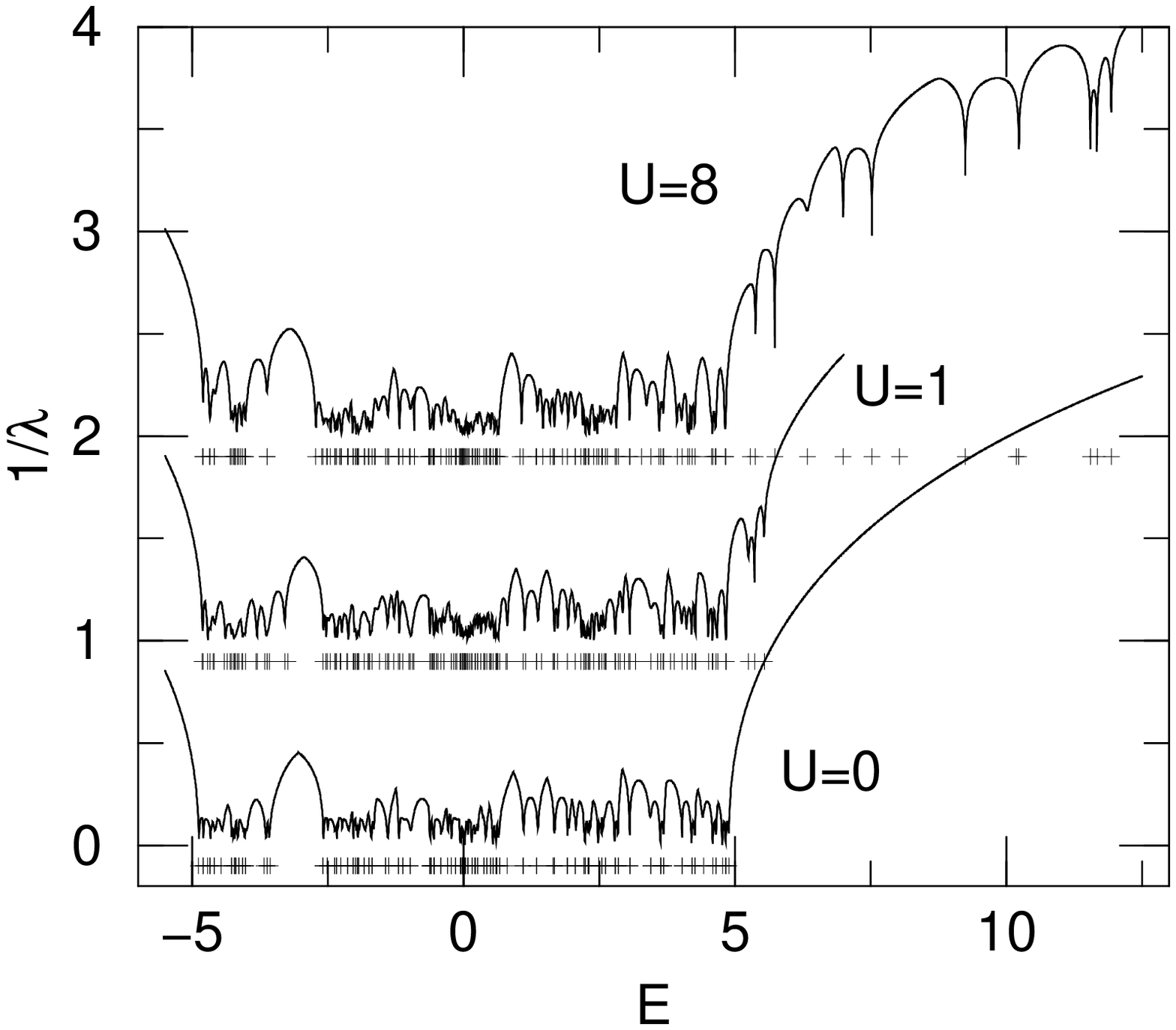,width=\figwidth}}
  \caption{
    Inverse localization length as a function of energy for a QP
    potential strength ${\mu} = 0.9$ at $\beta= \protect\sqrt{2}$,
    $M=13$ for $U=0$ and for two Hubbard interaction strengths $U$.
    Plots for different $U$ are vertically shifted for clarity. The
    eigenenergies are indicated by ($+$). }
\label{fig-EdepExt}
\end{figure}

\begin{figure}
  \centerline{\psfig{figure=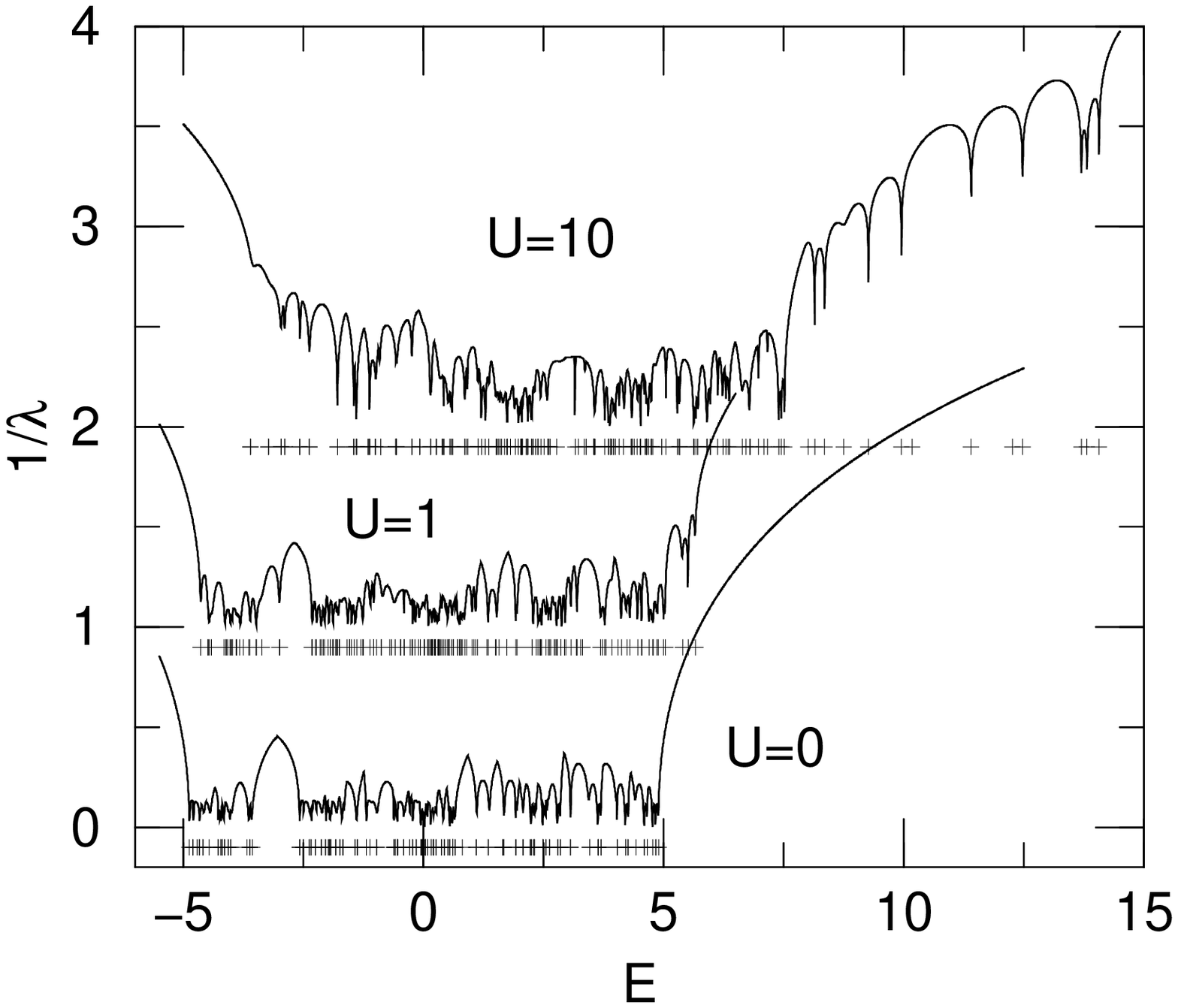,width=\figwidth}}
  \caption{
    Inverse localization length as a function of energy for a QP
    potential strength $\mu = 0.9$ at $\beta= \protect\sqrt{2}$,
    $M=13$ for $U=0$ and two long-range interaction strengths $U$.
    Plots for different $U$ are vertically shifted for clarity. The
    eigenenergies are indicated by ($+$). }
\label{fig-EdepExtLR}
\end{figure}


\begin{references}

\frenchspacing

\bibitem{AALR79} E. Abrahams, P. W. Anderson, D. C. Licciardello, and T. V. 
  Ramakrishnan, Phys. Rev. Lett. {\bf 42}, 673 (1979).

\bibitem{A58} P. W. Anderson, Phys. Rev. {\bf 109}, 1492 (1958).

\bibitem{BK94} D.\ Belitz and T. R. Kirkpatrick,
  Rev. Mod. Phys. {\bf 66}, 261 (1994).

\bibitem{KDS96} S. V. Kravchenko, D. Simonian, and M. P. Sarachik,
  Phys.  Rev. Lett. {\bf 77}, 4938 (1996);
%
  V. M. Pudalov, G.  Brunthaler, A. Prinz, and G. Bauer, preprint
  (1997, cond-mat/9707054); B. L. Altshuler and D. L. Maslov, preprint
  (1998, cond-mat/9807245); and references therein.

\bibitem{S94} D. L.\ Shepelyansky, Phys. Rev.  Lett. {\bf 73}, 2607
  (1994).

\bibitem{S96} D. L. Shepelyansky, in {\sl Correlated fermions and
    transport in mesoscopic systems}, Eds. T. Martin, G. Montambaux,
  and J. Tr\^{a}n Thanh V\^{a}n, Editions Frontieres,
  Gif-sur-Yvette, p. 201 (1996) (Proc. XXXI Moriond Workshop, 1996).

\bibitem{I95} Y. Imry, Europhys.  Lett. {\bf 30}, 405 (1995).

\bibitem{FMPW95} K. Frahm, A.  M\"{u}ller-Groeling, J. L. Pichard, and
  D. Weinmann, Europhys. Lett.  {\bf 31}, 169 (1995).

\bibitem{OWM96} F. v. Oppen, T. Wettig, and J. M\"uller, Phys. Rev.
  Lett.  {\bf 76}, 491 (1996).

\bibitem{WMPF95} D. Weinmann, A. M\"{u}ller-Groeling, J.-L. Pichard, and
  K. Frahm, Phys. Rev. Lett. {\bf 75}, 1598 (1995);
  T. Vojta, R. A. R\"{o}mer, and M. Schreiber, preprint (1997,
  cond-mat/9702241).

\bibitem{RS97} R.  A. R\"omer and M. Schreiber, Phys. Rev.  Lett. {\bf
    78}, 515 (1997); K. Frahm, A. M\"{u}ller-Groeling, J.-L.  Pichard,
  and D. Weinmann, Phys. Rev. Lett. {\bf 78}, 4889 (1997); R.  A.
  R\"omer and M. Schreiber, Phys. Rev.  Lett. {\bf 78}, 4890 (1997).

\bibitem{SK97} P. H. Song and D. Kim, Phys. Rev. B {\bf 56}, 12217
  (1997).

\bibitem{PS97} I. V. Ponomarev and P. G. Silvestrov, Phys. Rev. B {\bf
    56}, 3742 (1997).

\bibitem{CKM81} G. Czycholl, B. Kramer, and A.  MacKinnon, Z. Phys. B
  {\bf 43}, 5 (1981); J.-L. Pichard, J. Phys. C {\bf 19}, 1519 (1986).

\bibitem{BGKTMV98} D. Brinkmann, J. E. Golub, S. W. Koch, P. Thomas,
  K. Maschke, and I. Varga, preprint (1998).

\bibitem{LRS98} M. Leadbeater, R. A. R\"{o}mer, and M. Schreiber,
  preprint (1998, cond-mat/9806255); ---, preprint (1998,
  cond-mat/9806350).

\bibitem{HZKM98} O. Halfpap, A.  MacKinnon, and B. Kramer, to be
  published in Sol. State Comm., (1998); O. Halfpap, I. Kh.
  Zharekeshev, B. Kramer, and A.  MacKinnon, preprint (1998).

\bibitem{OS98} P. H. Song and F. v. Oppen, preprint (1998,
  cond-mat/9806303); K. Frahm, private communication.

\bibitem{AA80} S. Aubry and G. Andr\'e, Ann.\ Israel Phys.\ Soc.\ {\bf
    3}, 133 (1980).

\bibitem{H55} P. G. Harper, Proc. Phys. Soc. London Sect. A {\bf 68},
  874 (1955).

\bibitem{H76} D. R. Hofstadter, Phys. Rev. B {\bf 14}, 2239 (1976).

\bibitem{K83} M. Kohmoto, Phys. Rev. Lett. {\bf 51}, 1198 (1983); M.
  Kohmoto, L. P. Kadanoff, and C. Tang, Phys. Rev. Lett. {\bf 50},
  1870 (1983); S. Ostlund, R. Pandit, D. Rand, H. J. Schellnhuber, and
  E. D. Siggia, Phys. Rev. Lett. {\bf 50}, 1873 (1983); S. Ostlund and
  R. Pandit, Phys. Rev. B {\bf 29}, 1394 (1984); S. Das Sarma, S. He,
  and X. C. Xie, Phys. Rev. B {\bf 41}, 5544 (1990); I. Varga, J.
  Pipek, and B. Vasv\'{a}ri, Phys. Rev. B {\bf 46}, 4978 (1992).

\bibitem{BLT83} J. Bellissard, R. Lima, and D. Testard, Commun. Math.
  Phys. {\bf 88}, 207 (1983); B. Simon, Adv. Appl. Math. {\bf 3}, 463
  (1982).

\bibitem{BBJS96} A. Barelli, J. Bellisard, P. Jacquod, and D. L.
  Shepelyansky, Phys. Rev. Lett {\bf 77}, 4752 (1996), Phys. Rev. B
  {\bf 55}, 9524 (1997); D. L. Shepelyansky, Phys. Rev. B {\bf 54},
  14896 (1996).

\bibitem{ERS98} A. Eilmes, R. A. R\"{o}mer, and M. Schreiber, Eur.
  Phys. J.  B {\bf 1}, 29 (1998).

\bibitem{VES97} T. Vojta, F. Epperlein, and M. Schreiber, phys. stat.
  sol. (b) {\bf 205}, 53 (1998).

\bibitem{MK83} A. MacKinnon and B. Kramer, Z. Phys. B {\bf 53}, 1
  (1983).

\bibitem{evenodd} For $M$ an even Fibonacci number, we find that the
  SP localization lengths on the extended side $\mu <1$ show a
  qualitatively different behavior, shifting towards much larger
  localization lengths. This may be related to the fact that the QP
  potential $\mu_n \approx -\mu_{n+M/2}$ for all $n= 1, \ldots, M/2$.
  For TIP this shift is much less pronounced.

\bibitem{HJ85} R. A. Horn and C. R. Johnson, {\sl Matrix Analysis},
  Cambridge University Press, New York (1985).

\bibitem{GGS98} U.~Grimm, F.\ Gagel, and M.~Schreiber,
  in {\em Quasicrystals: Proceedings of the 6th Int.\ Conference},
  eds.\ S.~Takeuchi and T.~Fujiwara, p.\ 188 (World Scientific,
  Singapore, 1998).





\end{references}
\end{document}